\documentclass{article}

\usepackage{PRIMEarxiv}

\usepackage[utf8]{inputenc} 
\usepackage[T1]{fontenc}    
\usepackage{hyperref}       
\usepackage{cite}
\usepackage{amsmath,amssymb,amsfonts}
\usepackage{algorithm}
\usepackage{algpseudocode}
\usepackage{graphicx}
\usepackage{textcomp}
\usepackage{xcolor}
\usepackage{comment}
\usepackage{doi}
\usepackage{layout}
\usepackage{subfig}
\usepackage{siunitx}
\usepackage{placeins}
\usepackage{soul}
\usepackage{wrapfig}

\graphicspath{{figures}}    

\newcommand{\dirichlet}{\Gamma_\mathrm{D}}
\newcommand{\neumann}{\Gamma_\mathrm{N}}

\newcommand{\mbw}{\boldsymbol{w}}

\newcommand{\mbn}{\mathbf{n}}

\newcommand{\p}{p}

\newcommand{\mbnull}{\mathbf{0}}
\newcommand{\mbwd}{\boldsymbol{w}_\mathrm{D}}
\newcommand{\mbhr}{\hat{\mathbf{r}}}
\newcommand{\bx}{\boldsymbol{x}}
\newcommand{\ddx}[2]{\frac{\partial #1}{\partial #2}}
\newcommand{\cd}{c_\mathrm{D}}
\newcommand{\testFunc}{v}
\newcommand{\up}{c_{t}}
\newcommand{\un}{c_{t-1}}
\newcommand{\bc}{\mathbf{c}}
\newcommand{\cp}{\bc^{t}}
\newcommand{\cn}{\bc^{t-1}}
\newcommand{\dt}{\Delta t}
\newcommand{\bA}{\mathbf{A}}
\newcommand{\bL}{\mathbf{L}}
\newcommand{\dO}{\mathrm{d}\Omega}

\newcommand{\mun}{\mathbf{c}_{t-1}}
\newcommand{\win}{{w^\mathsf{in}}}
\newcommand{\wdi}{{w^\mathsf{dir}}}

\newcommand{\bfS}{{\mathbf S}}
\newcommand{\bfA}{{\mathbf A}}
\newcommand{\bfY}{{\mathbf Y}}
\newcommand{\bfC}{{\mathbf C}}
\newcommand{\bfH}{{\mathbf H}}
\newcommand{\bfO}{{\mathbf O}}
\newcommand{\bfR}{{\mathbf R}}
\newcommand{\bfT}{{\mathbf T}}
\newcommand{\bfV}{{\mathbf V}}
\newcommand{\bfd}{{\mathbf d}}
\newcommand{\bfP}{{\mathbf{P}}}
\newcommand{\bfE}{{\mathbf{E}}}
\newcommand{\bfI}{{\mathbf I}}
\newcommand{\bfw}{{\mathbf w}}
\newcommand{\bfc}{{\mathbf c}}
\newcommand{\bfeta}{{\boldsymbol{\eta}}}
\DeclareMathOperator*{\argmax}{arg\,max}

\newcommand{\clH}{\mathcal{H}}

\newcommand{\RR}{{\mathbb{R}}}

\newcommand{\beq}{\begin{equation}}
\newcommand{\eeq}{\end{equation}}
\newcommand{\bi}{\begin{itemize}}
\newcommand{\ei}{\end{itemize}}

\newcommand{\bF}{\mathsf{EnKF}}

\newcommand{\toCheck}[1]{\textcolor{black}{#1}}

\title{Sequential drone routing for data assimilation on a 2D airborne contaminant dispersion problem
}

\author{
  Daniele Giovanni Gioia,\textsuperscript{1}\thanks{Corresponding author}\,\,\,\,Jacopo Bonari,\textsuperscript{1} Daniel Lichte,\textsuperscript{1} Alexander Popp\textsuperscript{1,2} \\
 \textsuperscript{1}German Aerospace Center, Institute for the Protection of Terrestrial Infrastructures, Sankt Augustin, Germany.\\
 \textsuperscript{2}University of the Bundeswehr Munich, Institute for Mathematics and Computer-Based Simulation, Neubiberg, Germany
 \\
    e-mails:\texttt{\{daniele.gioia, jacopo.bonari, daniel.lichte\}@dlr.de, alexander.popp@unibw.de} \\
    ORCIDs:\{0000-0001-8979-4174, 0000-0001-8435-6466, 0000-0003-3314-5823, 0000-0002-8820-466X\}
}

\begin{document}
\maketitle

\begin{abstract}
The combined use of data from different sources can be critical in emergencies, where accurate models are needed to make real-time decisions, but high-fidelity representations and detailed information are simply unavailable.
This study presents a data assimilation framework based on an ensemble Kalman filter that sequentially exploits and improves an advection-diffusion model in a case study concerning an airborne contaminant dispersion problem over a complex two-dimensional domain.
An autonomous aerial drone is used to sequentially observe the actual contaminant concentration in a small fraction of the domain, orders of magnitude smaller than the total domain area. Such observations are synchronized with the data assimilation framework, iteratively adjusting the simulation. The path of the drone is sequentially optimized by balancing exploration and exploitation according to the available knowledge at each decision time.
Starting from an erroneous initial model based on approximated assumptions that represent the limited initial knowledge available during emergency scenarios, results show how the proposed framework sequentially improves its belief about the dispersion dynamics, thus providing a reliable contaminant concentration map.
\end{abstract}

\keywords{contaminant dispersion \and data assimilation \and dynamic routing \and sequential decisions}
\history{This is an Author’s Original Manuscript of an article presented at the 16th IEEE Sensor Data Fusion Symposium: Trends, Solutions and Applications in Bonn (DE) from 25-27 November 2024. Available online: \url{https://doi.org/10.1109/SDF63218.2024.10773899} Any citation must refer to its published version.}

\section{Introduction and related works}
During emergency situations, a rapid response that takes full advantage of all possible sources of information is essential to quickly proceed with critical decisions.
\toCheck{In the context of critical infrastructure protection}, this work investigates the dispersion of a single large burst of gas contaminant throughout a domain reproducing a real chemical factory. High-fidelity information about its dynamics are unavailable and data obtained from numerical simulations of an advection-diffusion model driven by air motion are coupled with in-situ sequential observations from an autonomous aerial drone observer at constant height, fused by means of a data assimilation strategy: the~\emph{Ensemble Kalman Filter} (EnKF).

Many explored solutions are purely based on models and/or simulations dependent on precise assumptions, both in terms of an accurate wind field evaluation in environments characterized by complex geometries~\cite{van_hooff_coupled_2010,blocken_computational_2015,tominaga_accuracy_2023} and of the analysis of the possible consequent atmospheric dispersion of hazardous materials~\cite{tominaga_cfd_2013,blocken_cfd_2013,toja-silva_urban_2018}.
However, the use of sensors, static and/or mobile, may reduce error\toCheck{s} and sequentially improve the knowledge of the underlying phenomenon by updating the current belief on the state of the system from a statistical perspective.


Data assimilation strategies fuse the statistical and dynamical information of limited ground-based observations and computational models, improving the knowledge about real physical phenomena. Most of the applications and algorithms come from geosciences~\cite{carrassi2018data}, but possible application domains and practical customized solutions are countless~\cite{asch2016data}. \toCheck{Several techniques come from the generalization and advancement of the well-known Kalman filter in control theory}. Variational methods are subject of research as well, typically more associated with weather forecasts, climate modeling, and oceanography. Still, hybrid methods are nowadays often the state of the art~\cite{asch2016data}. Bayesian statistics is usually the tool that allows for a sound exposition and interpretation of strategies coming from different domains. We refer to~\cite{asch2016data} for an overview of the applications and to~\cite{law2015data} for a mathematically rigorous introduction to data assimilation concepts.

Data assimilation techniques have also been succesfully applied in the context under examination. In~\cite{sharma2019estimating}, a finite set of static sensors is employed to track the contaminant distribution with unknown sources of an impulse release in two- and three-dimensional domains. The authors enhance a CFD contaminant transport model based on a probabilistic interpretation of the Perron-Frobenius operator and then integrate it within an EnKF. The use of \toCheck{an} EnKF for contaminant dispersion with static sensors is also implemented, e.g., by~\cite{lin2013forecasting}. Moving sensors are coupled with stationary ones in~\cite{darynova2023data} employing a methane-detecting drone, where a particle filter is selected as a data assimilation tool. A Gaussian plume dispersion model simplifies the transportation dynamics, defining a parametric state of the system where to apply Bayesian updates. However, the path where drones read the concentration is neither optimized nor sequentially adjusted according to the current belief about the state of the system. The use of drones on data-assimilation-based solutions for chemical transport problems has also been investigated on \toCheck{larger} domains \toCheck{(}i.e., entire regions\toCheck{)}, in~\cite{erraji2024potential} for air pollution predictions by using four dimensional variational methods on ozone and nitrogen oxide vertical profiles, coupling observations with the EURopean Air pollution Dispersion - Inverse Model (EURAD-IM). Similarly to~\cite{darynova2023data}, drones follow a not sequentially adjusted fixed path. 

In this work, we adaptively route an aerial vehicle to balance exploration and exploitation with regards to the current belief about the gas dynamics, whose model is provided by two sets of partial differential equations (PDEs). Their numerical solutions describe a wind vector field and the resultant atmospheric dispersion of the contaminant under examination, respectively. The steady state version of the \emph{incompressible Navier-Stokes} (INS) equations is employed to model the wind field, while the \emph{advection-diffusion} (AD) equation is used to simulate gas dispersion. For both sets of equations, the finite element method (FEM) is chosen as computational tool for their numerical solution.

Sequential routing of dynamic sensors with the aid of the (AD) equation, in the form of robotic swarms, is explored by~\cite{ wiedemann_experimental_2021,hinsen2023exploration} for the localization of the source of gas leakages. In~\cite{wiedemann_experimental_2021}, an ad-hoc method for automatic gas leakage localization based on Bayesian inference and named~\emph{Domain Knowledge Assisted Robotic Exploration and Source Localization} is experimentally validated with~\emph{static} assumptions and a domain without possible collision. On the other hand, in~\cite{hinsen2023exploration}, a Potential-Field-Control is utilized, such that devices move towards high uncertainty regions, similar to~\cite{wiedemann_experimental_2021}. However, their focus is antithetical to the presented work, as in this study the leakage source location is known, but gaining a high-fidelity dynamic contaminant concentration map throughout the zone of interest is assumed of critical importance. Furthermore, the use of a full Bayesian approach often poses harsh challenges in computational complexity and high dimensionality handling. Current experiments only consider small idealized domains with no obstacles (e.g.,~\cite{hinsen2023exploration,wiedemann_experimental_2021}). Data assimilation strategies, developed for high-dimensional problems with computationally expensive models, such as the EnKF, might allow for larger grid handling and, specifically, for sequential reconstruction of complex dynamics. Examples coupled with model predictive control, but only using the variance of the filters on simple domains are~\cite{ritter2014adaptive,euler2014centralized}. Conversely, we differently balance exploration and exploitation, working on a more complex domain and an ad hoc filter. Findings show how the assimilation framework dynamically provides valuable insight on the dispersion process.

\section{Airborne contaminant problem}

\subsection{Problem formulation}
The simulation framework considers a two-dimensional rectangular domain $\Omega$ bounded by user-defined linearized values of longitude $\lambda$ and latitude $\phi$. A representation set of real buildings is modeled inside $\Omega$ using their outer perimeter, resulting in a non-simply connected domain where the outer rectangular perimeter does not represent a proper physical boundary\toCheck{,} while the inner boundaries, coincident with buildings imprints, are considered obstacles for wind flow circulation, gas dispersion, and drone navigation. The simulation is performed on a time window $[0,T]$.

The domain boundary $\partial\Omega$ is divided into a Dirichlet boundary $\dirichlet$ and a Neumann boundary $\neumann$. On $\dirichlet$, a further split is performed. In correspondence of the outer rectangular boundary a wind vector field of constant intensity and direction is imposed, while on the inner boundaries, coincident with the buildings sides, a no-slip condition is enforced.

\begin{figure}
\centering
\includegraphics[width=0.7\columnwidth]{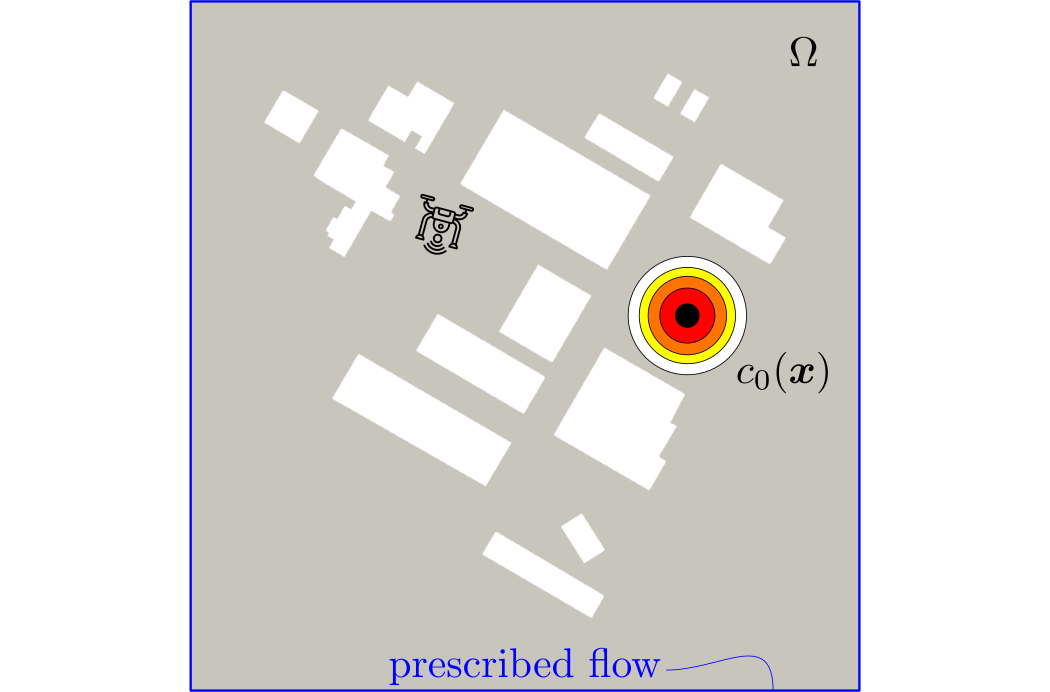}
\caption{Computational domain $\Omega$. A prescribed wind flow of constant direction and intensity is imposed \toCheck{on} the outer boundary for Eq.~\eqref{eq:INS_strong}. An initial gas concentration is defined as a scalar field $c_0(\boldsymbol{x})$, sketched together with a possible initial position of the drone at $t=0$.}
\label{fig:comp_dom}
\end{figure}

On $\Omega$, the wind flow is first evaluated by means of the constant viscosity convective form of the (INS) equations, then gas contaminant transport is simulated by means of the (AD) equation. Here, advection refers to the contaminant particles transport enforced by air motion, while diffusion describes the macroscopic averaged movement of particles resulting from their relative microscopic collisions. As a possible starting condition for the gas dispersion process, an initial distribution of contaminant, resulting for example from a burst or an abrupt leakage phenomenon, is prescribed in the form of a scaled Gaussian bell $c_0(\boldsymbol{x})$ with unitary maximum intensity and centered on a specific point of the domain. A sketch of the whole problem addressed is depicted in Fig.~\ref{fig:comp_dom}. The solution process implemented for the numerical solution of the~\eqref{eq:INS_strong} and the~\eqref{eq:ADstrong} equations is exhaustively exposed in the following paragraphs.

\subsubsection{INS Problem}
The steady state version of the (INS) equations in terms of kinematic viscosity $\nu$ and unitary density reads:
\begin{equation}\label{eq:INS_strong}
    \tag{INS}
    \begin{aligned}
        -\nu \nabla^2\mbw + \mbw\cdot\nabla\mbw + \nabla \p &= \mbnull \quad &\text{in} \quad &\Omega,\\
        \nabla \cdot \mbw &= 0 \quad &\text{in} \quad &\Omega, \\
        \mbw - \mbwd &= \mbnull \quad &\text{on} \quad &\dirichlet, \\
        \nu \partial \mbw/\partial \mbn - \p \mbn &= \mbnull \quad &\text{on} \quad &\neumann,
    \end{aligned}
\end{equation}
where $\mbw$ is the \toCheck{vector} field describing the wind velocity \toCheck{at} every point of $\Omega$ resulting from a value $\mbwd=\win\mbhr$ imposed on the outer boundary, \toCheck{with} $\win \in \RR_+$ \toCheck{being} the wind intensity and $\mbhr = [\cos(\wdi),\,\sin(\wdi)]^\intercal$, $\wdi \in [0,2\pi)$ the prescribed direction at the boundary. Finally, $\p$ is the \toCheck{scalar} intensity of the atmospheric pressure evaluated with respect to a reference point\toCheck{,} and $\mbn$ indicates the normal direction at each boundary point.

\toCheck{The finite element software FEniCS~\cite{alnaes2015fenics} is employed for the solution of Eq.~\eqref{eq:INS_strong}. Specifically, the equation is discretized in space using standard finite element procedures, which ultimately leads to a large system of nonlinear equations that has been addressed by a Newton-Raphson solver to obtain a solution in terms of the unknown discrete velocity and pressure degrees of freedom at every mesh node.}
\subsubsection{AD Problem}
The gas contaminant air transport phenomenon can be modeled by the following transient linear partial differential equation, that takes into account the two aforementioned contribution\toCheck{s} from advection and diffusion:
\begin{equation}
\tag{AD}
\mathcal{R}(c) \triangleq  \ddx{c}{t} + \mbw \cdot \nabla c - \varepsilon \, \Delta c, \label{eq:ADstrong}
\end{equation}
where $c$ is a scalar field describing the contaminant concentration, and $\varepsilon$ is the atmospheric diffusion coefficient for the agent under consideration. The complete initial boundary value problem thus reads:
\begin{align}\label{eq:IBVP}
\tag{IBVP}
    \begin{cases}
    	\mathcal{R}\bigl(c\bigl) = 0, &\;  \text{in} \; \Omega,  \\
    	c - c_0 = 0, & \;  \text{at} \; t=0,   \\
    	c - \cd = 0, & \;  \text{on} \; \dirichlet, \\
    	\partial c/\partial \mbn= 0, &\;  \text{on} \; \neumann. \\ 
    \end{cases}
\end{align}


The weak formulation of the~\eqref{eq:ADstrong} equation can be obtained by multiplying both members by a test function $\testFunc$ and applying the Green's identity to the diffusive term $\varepsilon\Delta c$:
\[
\label{eq:weakAD}
\tag{W-AD}
\int_\Omega \testFunc \cdot \frac{\partial c}{\partial t} \, \dO + 
\int_\Omega \testFunc \cdot \bigl( \mbw \cdot \nabla c \bigl) \, \dO +
\varepsilon \int_\Omega \nabla \testFunc \cdot \nabla c \, \dO = 0.
\]
\toCheck{The initial boundary value problem can as well be discretized in space within a finite element workflow, and in time by introducing an implicit backward Euler scheme}. The structure of the discretized problem can be expressed in terms of a bilinear form $a(\cdot)$, function of the field variable \toCheck{$\up$ at the current time step $t$} and of the test function $\testFunc$, and a linear form $L(\cdot)$ only dependent on the test function and on the (known) concentration at the previous time step \toCheck{$t-1$}:
\[
  a(\up,\testFunc) = L(\testFunc,\un).
  \label{eq:form}
  \tag{AD}
\]
After performing a standard FEM assembly process, the system state vector $\bc_t\in \RR_+^{M}$ at time $t$ that denotes the FEM solution of Eq.~\eqref{eq:form} in correspondence of the $M$ mesh nodes can be made explicit, and the final structure of the problem conveyed as a large system of linear algebraic equations, which reads: 
\[
    \bA \cdot \cp = \bL \cdot \cn.
    \label{eq:dt-AD}
    \tag{ADF}
\]
From this result, the state transition model can be implicitly expressed as: 
\beq \label{eq:transF} 
\tag{TransF}
\bA \cdot g(\mun|\text{(INS)}) = \bL\cdot\mun,
\eeq
where the system of equations is solved using a BiCGStab algorithm together with ILU preconditioning, with solution computed at $t-1$ used as initial guess at time $t$ to provide a warm-start to the iterative process.

As a final remark, for each analysis step, the values of velocity obtained from a previous time instant are employed in Eq.~\eqref{eq:weakAD}, and therefore considered to be still valid unless a change in the boundary conditions takes place, i.e., if variations from previous values of $\win$ and $\wdi$ do not fall within a specified threshold; in this case, the solution field coming from $t-1$ is used as a starting guess for a new Newton iteration to find an updated solution of the problem expressed by Eq.~\eqref{eq:INS_strong}, see Algorithm~\ref{alg:EnKF}.

\subsection{Domain generation}\label{subsec:dom_gen}
The domain generation, together with mesh discretization, are highly automated processes that leverage on the acquisition of the perimeter of real buildings through a query to OpenStreetMap~\cite{OpenStreetMap} and the GMSH~\cite{remacle_gmsh_2009} \toCheck{P}ython API Python-GMSH. The interested reader is referred to~\cite[\S3B]{bonari_towards_2024} for additional details on the geometry generation process. Specifically, the third out of three defined data acquisition paradigms has been employed, i.e., building data \toCheck{has} been prompted by means of a~\texttt{.geojson} file describing each building in terms of type and geometry. After that, a standard unstructured triangular mesh can be defined, considering the buildings as holes in the rectangular domain.

\section{Sequential data assimilation}
The sequential assimilation problem is formalized by following the sequential decision nomenclature of~\cite{Powell2022}. In particular, at a given time step $t$, assuming a FEM-based mesh with $M$ nodes, the knowledge about the contamination is summarized in a state of the system vector $S_t = [\bfc_t,\win_{\!\!\!t},\wdi_{\!\!\!\!\!t}\,
\,]$, including:
\bi
\item The concentration of the contaminant within the selected mesh structure $\bfc_t \in \RR_+^{M}$.
\item The wind intensity $\win \in \RR_+$ and direction $\wdi \in [0,2\pi)$, used to compute the velocity field 
in Eq.~\eqref{eq:INS_strong}.
\ei
Even if it is possible to explicitly consider other model quantities (e.g., the diffusion coefficient $\varepsilon$) as variables to be sequentially assimilated, they are not included in order to represent a real system, where the true model cannot be completely parameterized by a numerical simulation and only generates a low-fidelity twin. The sequentiality of the assimilation is indeed required to adjust concentration that would~\emph{not} be possible to simulate even if all the model parameters were perfect.

The assimilation system observes the concentration on a subset of nodes using an observation operator $\bfH_t \in \{0,1\}^M$, such that observation at time $t$ is $\bfO_t = \bfH_t(\bfc^{\mathsf{True}}_{t+1}) + \bfeta$, where $\bfc^{\mathsf{True}}_t$ is the true concentration projected onto the discretization and $\bfeta$ is an error term assumed independent and identically distributed Gaussian with zero mean and standard deviation $\sigma_{\mathsf{ob}}=10^{-3}$. This value corresponds to $0.1\%$ of the maximum of the normalized initial concentration (i.e., 1). It is worth noticing that the number of observed nodes per step is several orders of magnitude smaller than the number of nodes on the entire FEM-discretization. In this study, only one or two nodes per step are actually observed. Communication with the assimilation framework is assumed at $1.0\,\si{\Hz}$ frequency, synchronized with $\dt = 1.0\,\si{\second}$ in the time-discretized \eqref{eq:dt-AD}, together with a constant drone speed of $10.0\,\si{\metre\per\second}$. The wind intensity or direction is \emph{never} directly observed, but only implicitly filtered from concentration values.

\subsection{Ensemble Kalman filter}
The ensemble Kalman filter is a data assimilation method based on Gaussian hypotheses and running an ensemble of $N$ \emph{trajectories}, i.e., a set of different model dynamics based on a Monte Carlo sampling. For an extended discussion about the various versions and a deeper explanation, we refer to~\cite{asch2016data}. The implemented version is an \emph{Ensemble Transform Kalman Filter} (ETKF) that exploits the eigenvalue decomposition of Hermitian matrices and is inspired by the implementation on DAPPER~\cite{raanes2024dapper}. It currently does not include either inflation or localization. Future works will be devoted to adapting such tools to the case study.

To create the ensemble at $t=0$, wind directions are sampled according to the initial belief of the model. Specifically, the initial value of intensity $\win_{\!\!\!\!\mathsf{init}}$ and direction $\wdi_{\!\!\!\!\!\!\mathsf{init}}$ are defined according to the belief the system has \emph{before} any observation, thus representing the initial knowledge about the real phenomenon (e.g., from weather forecasts). The ensemble samples for each trajectory $i \in \{1,\dots,N\}$:
\[
\win_{\!\!\!0,i} \sim \text{Weibull}(\win_{\!\!\!\!\mathsf{init}},\sigma_\mathsf{in}) \quad \forall i \in \{1,\dots,N\},
\] 
for the wind intensity, parametrizing the Weibull distribution by using the mean $\win_{\!\!\!\!\mathsf{init}}$ and standard deviation $\sigma_\mathsf{in}$. Regarding the wind direction, it is assumed to follow a modular normal distribution as:
\[
\wdi_{\!\!\!\!\!0,i} \sim \left[\mathcal{N}(\wdi_{\!\!\!\!\!\!\mathsf{init}}, \sigma_\mathsf{dir}) \right]_{\mathsf{mod} \,2\pi}  \quad \forall i \in \{1,\dots,N\},
\]
to consider the circular properties of the angles. Within the assimilation framework, wind values in terms of intensity and direction are employed as boundary conditions for the problem defined by Eq.~\eqref{eq:INS_strong}, generating different trajectories with different velocity fields. Given an ensemble made of $N$ trajectories, for each time step $t$, the state of the system is redefined as:
\[
\bfS_t = [S_t^1,\dots,S_t^N],
\]
therefore representing the current belief by means of the current ensemble values. 
Equation~\eqref{eq:transF} provides the transition step for each ensemble trajectory. The belief about the wind parameters is only updated during assimilation.

Given the current state of the system, it is easier to consider a matrix form that reads
\[
\bfS_t = [\bfC_t,\bfw^\mathsf{I}_t,\bfw^\mathsf{d}_t],
\]
where 
\begin{align*}
\bfC_t &= [\bfc^1_t, \dots,\bfc^N_t]^{\mathsf{T}} \in \RR^{N\times M}_+,\\
\bfw^\mathsf{I}_t &= [\win_{\!\!\!t,1},\dots,\win_{\!\!\!t,N}]^{\mathsf{T}} \in \RR_+^N,\\
\bfw^\mathsf{d}_t &= [\wdi_{\!\!\!\!\!t,1},\dots,\wdi_{\!\!\!\!\!t,N}]^{\mathsf{T}} \in [0,2\pi)^N.
\end{align*}
Moreover, it is useful to define the post-dynamics pre-assimilation state 
\[
\bfS^\dag_t = [g(\bfS_t),\bfw^\mathsf{I}_t,\bfw^\mathsf{d}_t] =[\bfC^\dag_t,\bfw^\mathsf{I}_t,\bfw^\mathsf{d}_t],
\]
where $g$ comes from Eq.~\eqref{eq:transF}, only affecting the concentration, and the new post-assimilation belief is
\[
\bfS_{t+1} = \bF(\bfS^\dagger_t, \bfO_t)=[\bfC_{t+1},\bfw^\mathsf{I}_{t+1},\bfw^\mathsf{d}_{t+1}].
\]
The complete algorithm is summarized in Algorithm~\ref{alg:EnKF}. Notice that, due to the polar nature of the wind characteristics, an ad-hoc handling of the direction assimilation is required. Moreover, $\mu$ is the mean with regard to the ensemble and operations are meant in a vectorized sense. Wind values $(\bfw^\mathsf{I}_t)^\text{(INS)}_i$ and $(\bfw^\mathsf{d}_t)^\text{(INS)}_i$ are the last values of wind intensity and direction for which \eqref{eq:INS_strong} was solved for each trajectory~$i$. Within the model, noise $\boldsymbol{\Sigma}$ is sampled from a centered independent Gaussian distribution with covariance matrix with standard deviation equal to 10\% of the correspondent trajectory concentration value and included \emph{before} the transition step. To limit the (INS) re-computations in this example we do not add direct noise to $\bfw^\mathsf{I}_t,\bfw^\mathsf{d}_t$. However, the algorithm can include it.


\begin{algorithm}[tb]
\footnotesize
\caption{Complete EnKF step ( \textbf{Input:} $\bfS_t$; \textbf{Output:} $\bfS_{t+1}$)}\label{alg:EnKF}
\begin{algorithmic}
\State \State  $\clH_t^\pi(\bfS_t) = \bfH_t$ \Comment{Define the observation according to \eqref{eq:pol}}
\State
\State $\bfC_t = (\bfC_t + \boldsymbol{\Sigma})\cdot\boldsymbol{1}_{(\bfC_t + \boldsymbol{\Sigma}>\epsilon)}$  \Comment{Inter step additive noise on concentration if non spurious}\vskip 0.9pt
\State
\For{$i \in \{1,\dots,N\}$} \Comment{Re-solve \eqref{eq:INS_strong} if necessary}
\If{$ (|(\bfw^\mathsf{I}_t)_i^ - (\bfw^\mathsf{I}_t)^\text{(INS)}_i| \ge \mathsf{tr}_{\mathsf{I}}\cdot |(\bfw^\mathsf{I}_t)_i|)$ OR $ (|(\bfw^\mathsf{d}_t)_i - (\bfw^\mathsf{d}_t)^\text{(INS)}_i| \ge \mathsf{tr}_{\mathsf{d}}\cdot |(\bfw^\mathsf{d}_t)_i|)$} \vskip 0.9pt
\State Solve \eqref{eq:INS_strong} with $(\bfw^\mathsf{I}_t)_i$ and  $(\bfw^\mathsf{d}_t)_i$
\EndIf
\EndFor
\State
\State $\bfC_t^\dag = g(\bfS_t) = g(\bfC_t|\text{(INS)})$ \Comment{AD-dynamical step}
\State
\State $\bfA_t = \bfC_t^\dag - \mu(\bfC_t^\dag)$  \Comment{Centered anomaly matrix for concentration}
\State $\bfA^\mathsf{I}_t = \bfw^\mathsf{I}_{t} - \mu(\bfw^\mathsf{I}_{t} )$ \Comment{Centered anomaly matrix for wind intensity}
\State $\bfA^\mathsf{d}_t,\, \mu_{\mathsf{circ}} = \text{CIRC\_AN}(\bfw^{\mathsf{d}}_{t})$ \Comment{Centered anomaly matrix for wind direction and circular mean}
\State
\State $\bfO_t = \bfH_t(\bfc^{\mathsf{True}}_{t+1}) + \bfeta$ \Comment{Observe}\vskip 0.9pt
\State $\bfY = \bfH_t(\bfC_t^\dag) - \mu(\bfH_t(\bfC_t^\dag))$ \Comment{Ens. observation anomalies}\vskip 0.9pt
\State $\mathsf{dy} =\bfO_t  -  \mu(\bfH_t(\bfC_t^\dag)) $ \Comment{Mean innovation w.r.t. current belief}
\State
\State $\bfR^{-1} = \text{Diag}(\frac{1}{\sigma_\mathsf{ob}})$ \Comment{Inverse observation error diagonal matrix}\vskip 0.9pt
\State $\bfd,\,\bfV = \text{EIG}(\bfY \bfR^{-1} \bfY^{\mathsf{T}} + (N-1)\bfI_{N} )$ \Comment{Eigval, eigvect ($\bfI_{N}$: identity)}\vskip 0.9pt
\State $\bfT = \sqrt{N-1} \cdot \bfV \, \text{Diag}(\frac{1}{\sqrt{\bfd}}) \bfV^\mathsf{T}$\vskip 0.9pt
\State $\bfP = \bfV \text{Diag}(\frac{1}{\bfd}) \bfV^\mathsf{T}$\vskip 0.9pt
\State $\Lambda = (\mathsf{dy}\, \bfR^{-1}) (\bfY^{\mathsf{T}} \bfP)$ \vskip 0.9pt
\State $\bfE = [\mu(\bfC_t^\dag), \mu(\bfw^\mathsf{I}_{t}), \mu_{\mathsf{circ}}] + \Lambda [\bfA_t,\bfA^\mathsf{I}_t, \bfA^\mathsf{d}_t] + \bfT [\bfA_t,\bfA^\mathsf{I}_t, \bfA^\mathsf{d}_t]$\vskip 0.9pt
\State $[\bfC_{t+1}]_{ij} = \max[0, \bfE_{ij}] \quad \forall i \in \{1,\dots,N\} \, \forall j \in \{1,\dots,M\}$\vskip 0.9pt
\State $(\bfw^\mathsf{I}_{t+1})_i = \max[0,\bfE_{i,M+1}] \quad \forall i \in \{1,\dots,N\}$\vskip 0.9pt
\State $(\bfw^\mathsf{d}_{t+1})_i = [\bfE_{i,M+2}]_{\text{mod} 2\pi} \quad \forall i \in \{1,\dots,N\}$
\State
$\bfS_{t+1} = [\bfC_{t+1},\bfw^\mathsf{I}_{t+1},\bfw^\mathsf{d}_{t+1}]$
\State

\Function{circ\_an}{$\bfw^\mathsf{d}_{t}$}
\State $\mu_{\sin} = \mu(\sin{(\bfw^\mathsf{d}_{t})})$  \Comment{mean of sin of the trajectory directions}
\State $\mu_{\cos} = \mu(\cos{(\bfw^\mathsf{d}_{t})})$  \Comment{mean of cos of the trajectory directions}
\State $\mu_{\mathsf{circ}} = [\text{arctan2}(\mu_{\sin},\mu_{\cos})]_{\text{mod} 2\pi}$
\State $\bfA^\mathsf{d}_t = \bfw^\mathsf{d}_{t} - \mu_{\mathsf{circ}}$
\State $\bfA^\mathsf{d}_t = [\bfA^\mathsf{d}_t + \pi]_{\text{mod} 2\pi} - \pi $\Comment{anomalies in $[-\pi,\pi)$}
\State return $\bfA^\mathsf{d}_t,\, \mu_{\mathsf{circ}}$
\EndFunction

\end{algorithmic}
\end{algorithm}

\subsection{Cost Function Approximation routing}

Given the state of the system $\bfS_t$, the next points to be observed are chosen according to a policy $\pi$. Since the selection problem is inherently sequential and policy-dependent, it is possible to explicitly formalize the observation operator as the outcome of a decision policy, such that
\beq\label{eq:pol}\tag{D}
\clH_t^\pi: \bfS_t \to \{0,1\}^M
\eeq
and $\clH_t^\pi(\bfS_t) = \bfH_t$ is an observation operator such that $\bfO_t = \bfH_t(\bfc^{\mathsf{True}}_{t+1}) + \bfeta$. Such a policy is implicitly subject to the physical constraints of the observer (i.e., the drone cannot teleport and can only move a certain distance between two time steps) and follows the space discretization structure of the mesh when the information is synchronized with the simulated model. 

Because of the large domain $\Omega$, the drone has a maximum movement per time-step orders of magnitude smaller. For this reason, rather than deciding the sequential path node by node, we select the direction of the next movement, then project the direction on the mesh, retrieving the correspondent nodes that are visited during the in-step time interval, therefore generating the observation operator embedded within the mesh. In case of additional static observers (i.e., fixed sensors like in~\cite{darynova2023data}), the operators would simply include their respective closest nodes by default, also synchronizing, where necessary, different reading time intervals.

We introduce a policy that consists of a \emph{Cost Function Approximation} (CFA) approach~\cite{Powell2022}. Since our initial goal would be to observe where the uncertainty is greatest, we use the cumulative standard deviation of the anomalies matrix to \emph{exploit} according to the current belief and the cumulative concentration mean to \emph{explore} it, entailing an approximate cost. Therefore, the drone is dynamically routed according to an exploration-exploitation balance driven by the inherited uncertainty of the ensemble sampling.

Given the current position of the device in 2D Cartesian coordinates, $\bx_t = (x^o_t, y^o_t)$, we divide the domain $\Omega$ in $Q$ circular sectors by using as origin $\bx_t$. Specifically, $Q$ subsets of nodes are defined as:
\begin{align*}
    \bfC^{q}_t = \bigg\{ [\bfC_t]_{ij}:& [\text{arctan2}(y^j_t-y^o_t, x^j_t -x^o_t)]_{\text{mod} 2\pi}  \in [(q-1)\frac{2\pi}{Q},q\frac{2\pi}{Q} ),
    \\\,&\forall q \in \{1,
\dots,Q\} \,\forall i \in \{1,
\dots,N \} \,\forall j \in \{1,
\dots,M\} \bigg\},
\end{align*}
where coordinates $(x^j_t,y^j_t)$ identify nodes of the mesh with $N$ different concentration evaluations according to ensemble trajectories. $M^{q}$ is the number of nodes per subset $q$. 
A direction is determined by the angle:
\[
q \to q\frac{2\pi}{Q} - \frac{\pi}{Q},
\]
mid-angle of each sector $q$. 
The direction is selected by computing, for each $q$, the CFA and selecting the highest value. To compute the cumulative standard deviation, the trace of the covariance matrix along the ensemble axis is calculated. Its square root is computed and weighted with the sum of the mean concentration in the sector according to a parameter $\theta_t$. More explicitly, considering a sector $q$, the subset anomaly matrix (not considering wind direction and intensity) is
\[
\bfA_t^{q} = \bfC^{q}_t - \mu(\bfC^{q}_t),
\]
where $\mu(\bfC^{q}_t) \in \RR_+^{M^q}$.

The CFA policy selects the direction by maximizing
\beq\label{eq:CFA}\tag{EnIE}
 \!\!\argmax_{ q \in \{1,\dots,Q\}} \left\{\!\!\frac{1}{M^q}\bigg( \sqrt{\mathsf{tr}\bigg(\frac{\bfA_t^{q}{\bfA_t^{q}}^\mathsf{T}}{N-1}\bigg)}\!+\theta_t\!\!\sum_{j=1}^{M^q} \mu(\bfC^{q}_t)_j \!\bigg)\!\!\!\right\},
\eeq
where $\theta_t$ can be optimized according to specific objective functions. It is also possible to partially overlap circular sectors for high values of $Q$ and/or to directly evaluate a more complex path with different subsets that take into account the characteristics of the specific problem. Nevertheless, it is important to generate a non-myopic policy that considers the value of the current decision in the future. We refer to~\cite{Powell2022} for an extensive discussion of sequential optimization problems. The introduced policy is defined as \emph{Ensemble Interval Estimation} (EnIE-2D). The three-dimensional policy generalization is a future work and should intuitively be based on spherical sectors.

When the selected direction is physically blocked by an obstacle, the closest free angle is alternatively selected. In particular, $-2\pi/Q$ are iteratively added to generate a clockwise alternative until a viable option with no obstacle is found. It is worth noticing that, even if the policy is non-myopic, the handling of obstacles is currently na\"{i}ve and improvements will follow. Currently, a low value of $Q$, especially with non-convex buildings, might heavily prolong the path, temporarily locking the device in the case of peculiar configurations.

\section{Computational experiments}

\subsection{Design of experiments}
To evaluate the performance of the assimilation framework, we run a twin model where one simulation is used as reference, therefore representing the truth. Its parameters are set as $\win_{\!\mathsf{True}} = 5.0\,\si{\metre\per\second}$, $\wdi_{\!\!\!\mathsf{True}} = \pi$ and $\varepsilon = 0.8\,\si{\metre^2\per\second}$. As initial belief, incorrect parameters are assumed, thus defining an inaccurate initial condition that we aim to sequentially improve, chasing the correct dynamics. Specifically, we choose $\win_{\!\!\!\!\mathsf{init}} = 2.5\,\si{\metre\per\second}$, $\wdi_{\!\!\!\!\!\!\mathsf{init}} = 3\pi/2$ and $\varepsilon = 0.4\,\si{\metre^2\per\second}$. It is important to remark that $\varepsilon$ is inaccurate and cannot be learned, i.e., it represents parameters that are not possible to model when dealing with reality rather than a twin model experiment. The ensemble is set up with $N=10$ trajectories sampled with $\sigma_\mathsf{in} = 3.0\,\si{\metre\per\second}$ and $\sigma_\mathsf{dir}=\pi/2$. Moreover, $\mathsf{tr}_{\mathsf{I}}=10\%$, $\mathsf{tr}_{\mathsf{d}}=5\%$, and $Q=8$. The simulation lasts $T=100.0\,\si{\second}$ with a grid having $M\approx 10^4$ degrees of freedom. The domain is generated according to the procedure exposed in Sec.~\ref{subsec:dom_gen}, with building data obtained from a \texttt{.geojson} file reproducing the building layout of an existing chemical plant. The assimilation frequency is assumed to be $1\,\si{\Hz}$. However, thanks to the current offline twin-experiment setting, if some internal simulations happen to be longer, they do not provoke a de-synchronization. The number of trajectories and the solution time for Eq.~\eqref{eq:INS_strong} play an important role regarding the practical assimilation frequency. An ad-hoc tuning is definitely required.

The true concentration $\bfc^\mathsf{True}_t$ is first compared to a model run without any assimilation, i.e., $\bfc^{\mathsf{init}}_t$, to assess the inaccuracy effect. Sequential assimilation is tested using policy~\eqref{eq:CFA} with three different choices for $\theta_t$:
\bi
\item $\bfc^{\mathsf{EnIE_0}}_t$; where $\theta_t = 0.0 \,\,\, \forall t \in \{0,\dots,T\}$ and only the cumulative standard deviation is employed, generating a purely exploitative path.
\item $\bfc^{\mathsf{EnIE_{0.2}}}_t$; where $\theta_t = 0.2 \,\,\, \forall t \in \{0,\dots,T\}$.
\item  $\bfc^{\mathsf{EnIE_{0.3}}}_t$; where $\theta_t = 0.3 \,\,\, \forall t \in \{0,\dots,T\}$.
\ei

\begin{figure}
\centering
\includegraphics[width=0.5\columnwidth]{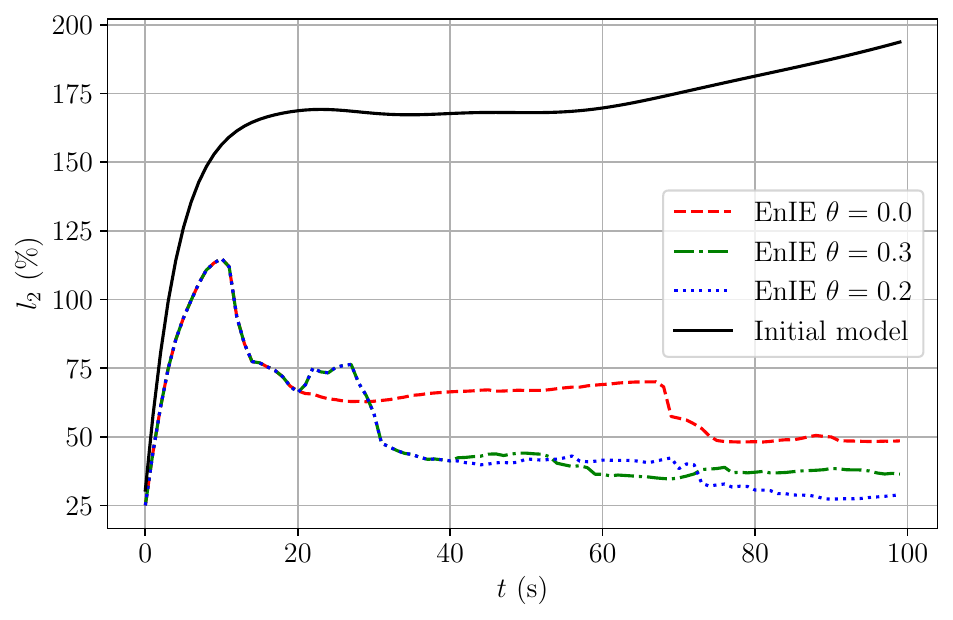}
\caption{Quantitative error analysis for different $\theta$ values and the initial model without assimilation.}
\label{fig:err_plots}
\end{figure}

\subsection{Evaluation metrics and results}
Given a concentration value $\bfc^*_t$, the relative $l_2$-difference at time $t$ is computed as:
\[
\text{ERR}_{l_2}=\frac{\|\bfc^\mathsf{True}_t - \bfc^{*}_t\|_2}{\|\bfc^\mathsf{True}_t\|_2} \cdot 100.
\]
Results are summarized in Fig.~\ref{fig:err_plots}, where it is possible to notice how the assimilation effect initiates after an initial transient phase used by the drone to get closer to the contaminant. As soon as the drone is close enough, different balance parameters generate different results. The best policy for the presented case study is $\theta=0.2$. However, the path dependence on the parameter is nontrivial. An explicitly time-dependent $\theta_t$ is currently subject of research. Regardless of $\theta$, the improvement of the quality of the dynamics achieved by the presented assimilation framework becomes obvious, since the initial model alone would completely diverge from reality.

\begin{figure}
\centering
\includegraphics[width=0.5\textwidth]{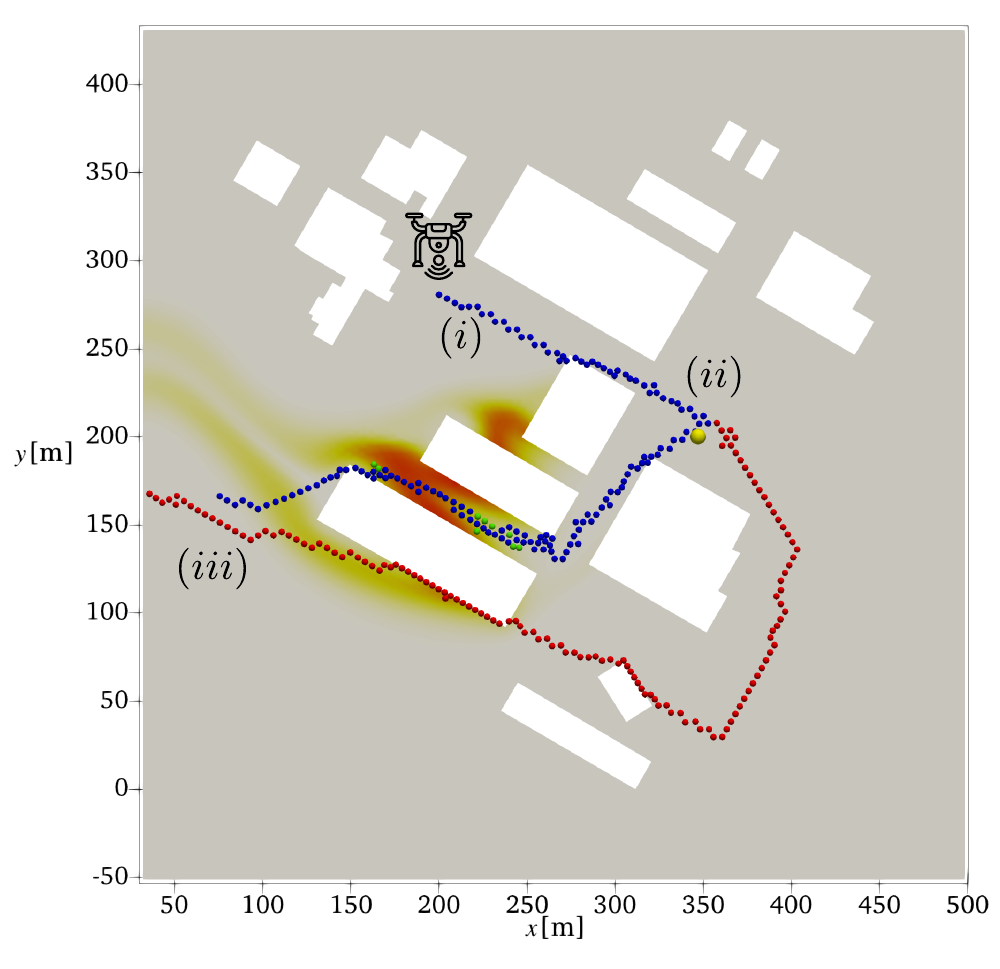}
\caption{Nodes of the mesh visited for  $\theta = 0.0$ (red); $\theta = 0.3$ (green) and $\theta = 0.2$ (blue). Red points often overlap with blue ones. Concentration field at $t = T$ as background. Drone initial position $(i)$, initial release location marked by a yellow circular glyph $(ii)$ and drone final position at $t=T$ $(iii)$.}
\label{fig:traj_plots}
\end{figure}

A better understanding of the influence of the parameter $\theta$ comes from Fig.~\ref{fig:traj_plots}, where the drone path is presented for different values of $\theta$. The mean concentration of the belief at $t=100.0\,\si{\second}$ is given as background contour plot together with the initial position of the drone and the release location. The more the policy explores, the sooner it steers towards the new high-concentration zone according to its belief as it gets closer to the release. High exploitation policy ($\theta = 0.0$) stay in the release zone longer due to the higher variance of the trajectories, then following a path where the concentration should have been, but where it is not anymore, thus being the location where the belief is suddenly shifted the most.

\par

\begin{figure*}[!b]
  \centering
  \includegraphics[width=0.8\textwidth]{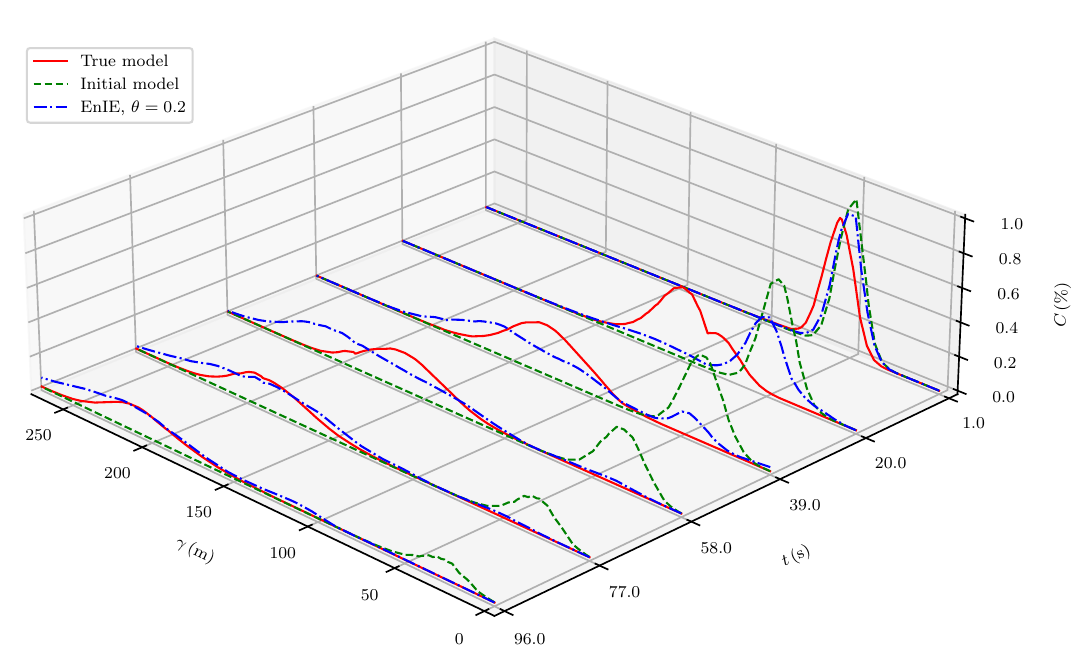}
  \caption{Concentration values over the curve $\gamma(s)$. True model, model based only on the initial assumptions, and filter with $\theta = 0.2$, respectively.}
  \label{fig:conc_plot}
\end{figure*}

\begin{figure*}[!ht]
	\centering
    \subfloat[][True model, $t = 0.0$.\label{subfig:}]
	{\includegraphics[width=0.45\columnwidth]{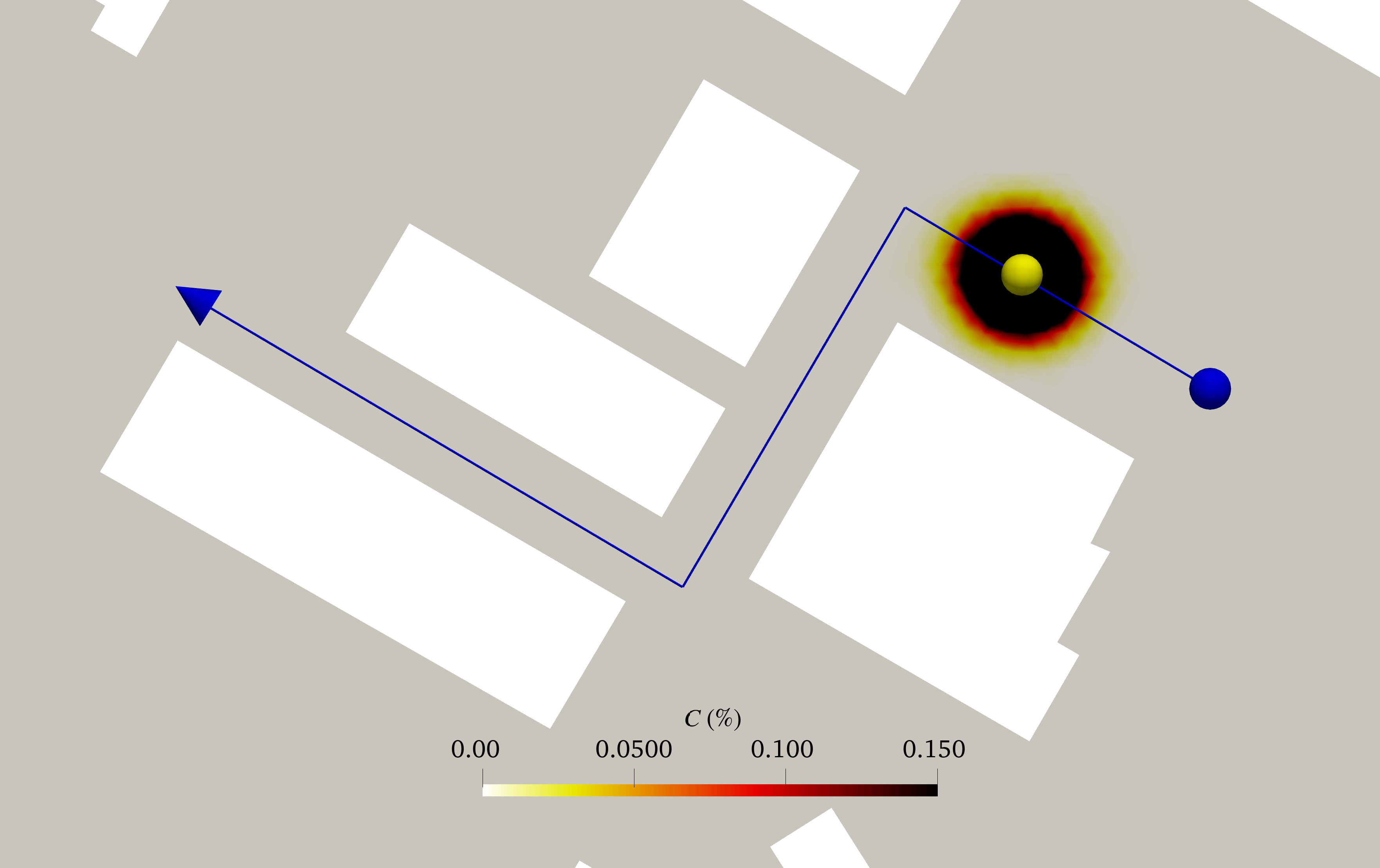}}
	\hspace{5pt}
    \subfloat[][True model, $t = T$.\label{subfig:true_contour_4b}]
	{\includegraphics[width=0.45\columnwidth]{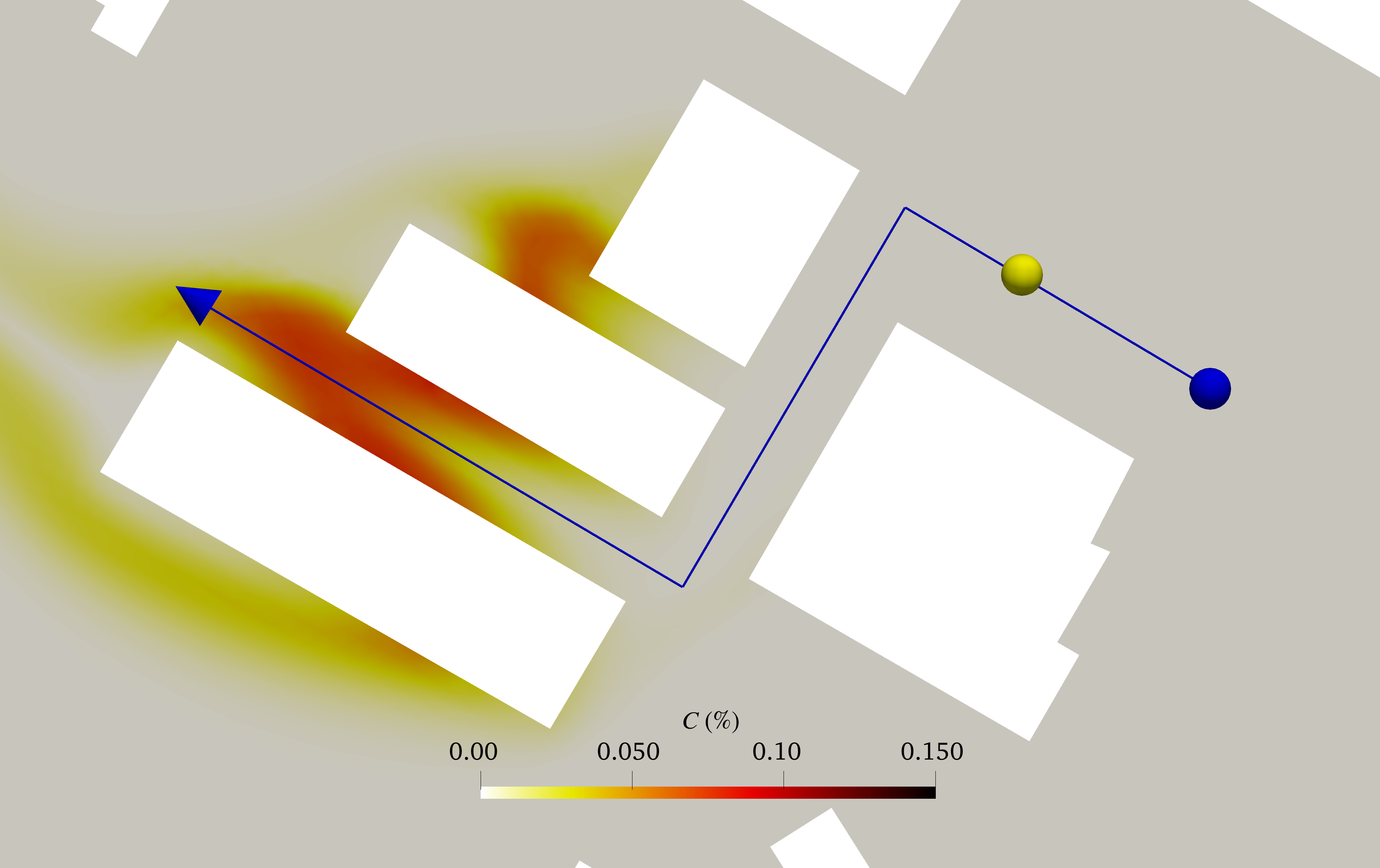}}\\
	\subfloat[][Model run with initial assumptions alone, $t = T$.\label{subfig:bias_model}]
	{\includegraphics[width=0.45\columnwidth]{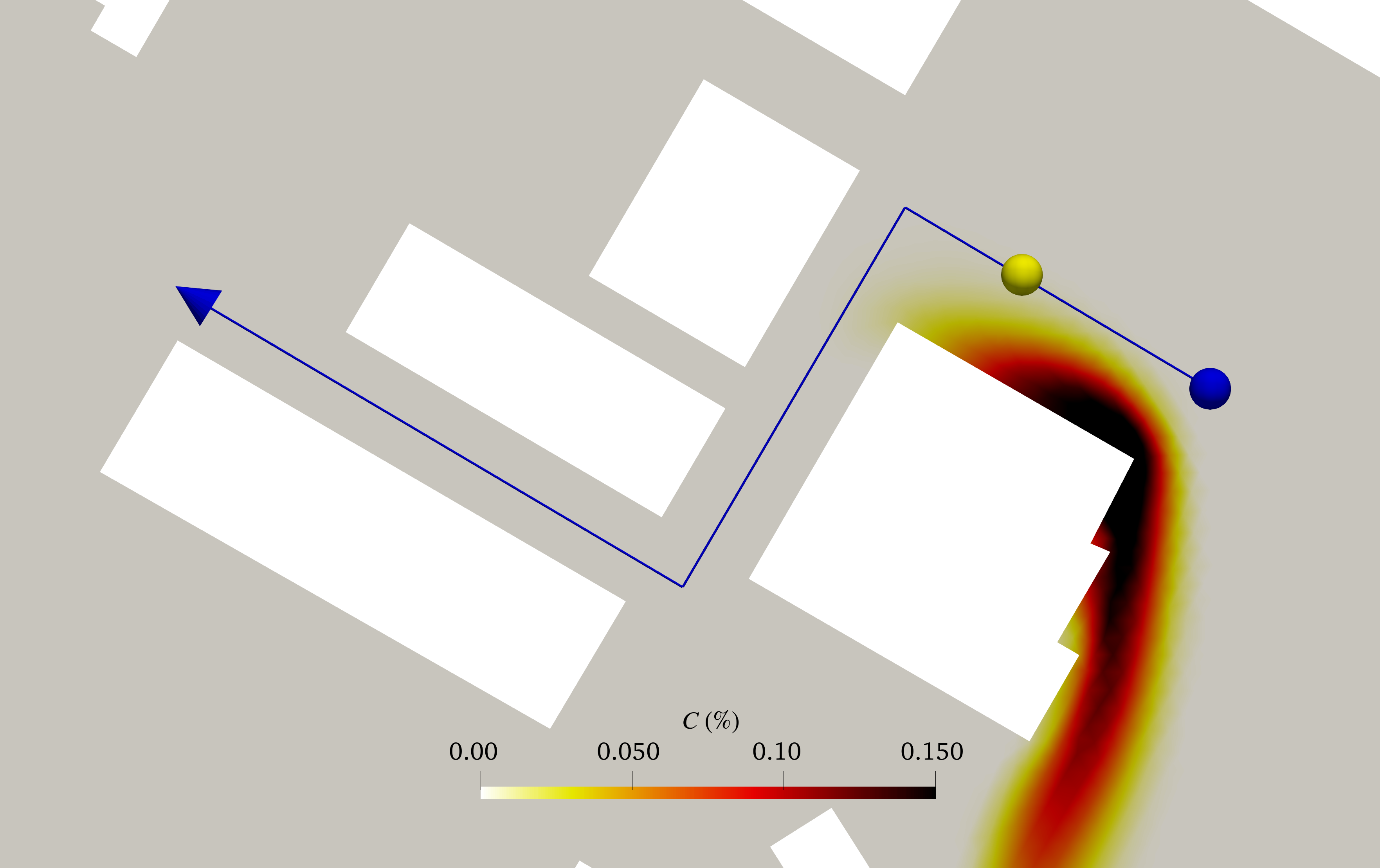}}
    \hspace{5pt}
    \subfloat[][Filtered model, $t = T$.\label{subfig:filter_model}]
	{\includegraphics[width=0.45\columnwidth]{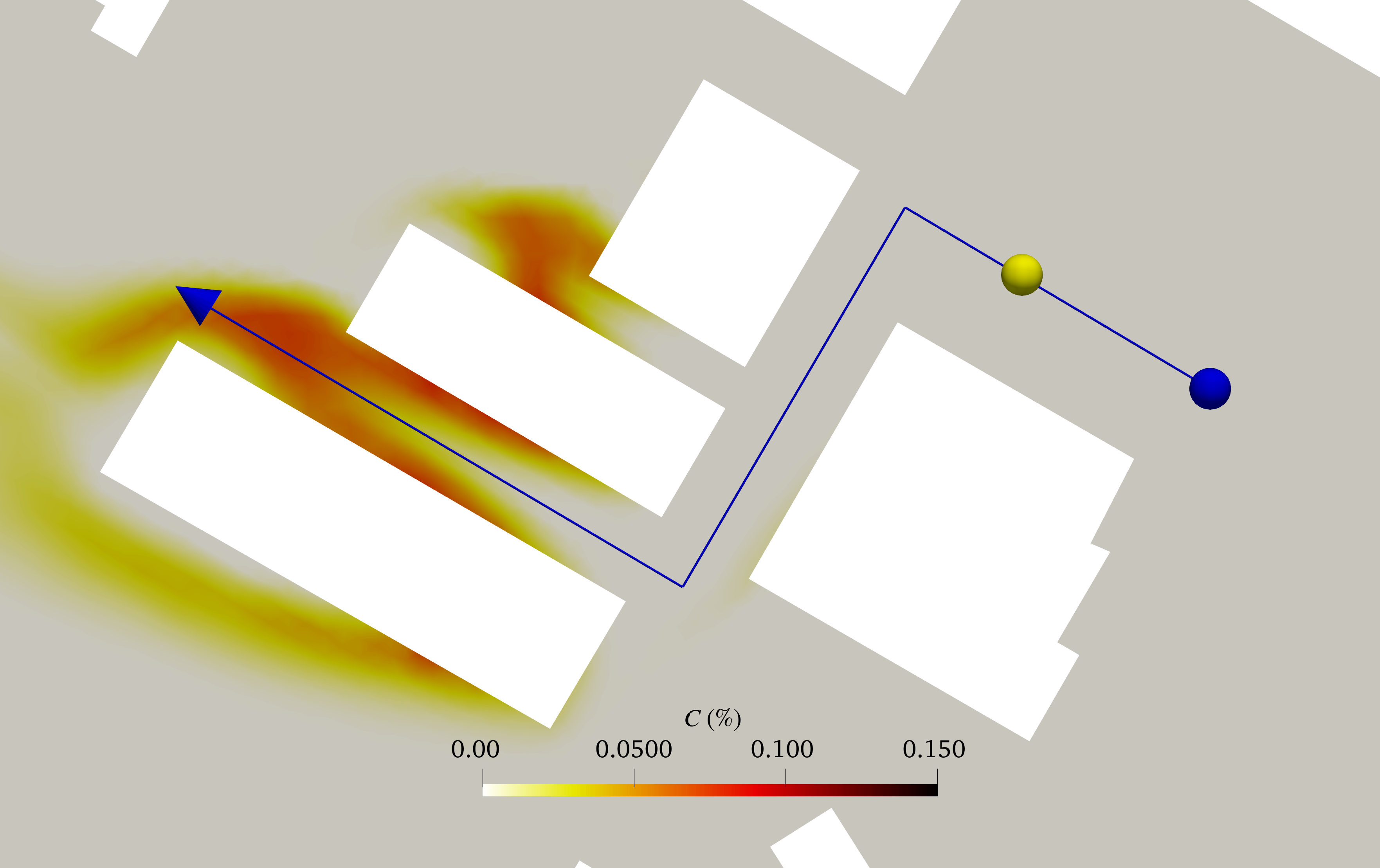}}
	\caption{Four representative contour plots for the contaminant concentration: the starting condition at  $t = 0.0$, the reference true model  at $t = T~(100.0\,\si{\second})$, the model based only on initial conditions, at the same time step, and the model reconstructed through data assimilation with $\theta=0.2$, still at $t = T$. A curve $\gamma(s)$ is highlighted by the blue poly-line and employed as section cut for a temporal evaluation of the contaminant field in Fig.~\ref{fig:conc_plot}.}
	\label{fig:rep_plots}
\end{figure*}

To generate a temporal overview of the entire assimilation process throughout the selected horizon, Fig.~\ref{fig:conc_plot} plots the concentration over time on a curve of interest $\gamma(s)$ identified in  Fig.~\ref{fig:rep_plots} and representative of the assimilation process. The filter effectively improves the quality of its belief about the concentration dynamics after $t=20.0\,\si{\second}$, around the time of arrival to the neighborhood of the release location. Afterward, the filter gets closer and closer to the true dynamics, abandoning the model run with an incorrect initial belief. Figure~\ref{fig:rep_plots} also presents the concentration at time $T$ according to the true model, the inaccurate model relying only on the initial assumptions, and the filter. The dynamics of the filter resemble the ones of the true model, while being far from the inaccurate model.

\section{Conclusions}
To cope with emergencies in the context of critical infrastructures protection, the use of all available data, both synthetic and observed in the field, is critical. In this study, a data assimilation framework for a contaminant transport problem was presented, based on an ensemble Kalman filter and an advection-diffusion model. Routing policies were introduced for a drone that observes the actual concentration in the field and fuses the information with a dynamic model. The results showed that even a small number of observations and a simple balancing of exploration and exploitation against the knowledge available to the data assimilation framework can generate a system capable of adaptively adjusting the simulation, thus creating a valid decision-making tool to be employed in, e.g., evacuation scenarios. Experiments were performed on a complex, yet two-dimensional domain. Future work will involve generalization to a three-dimensional domain, better management of the drone's physical constraints, and improvement of the data assimilation algorithm.

\FloatBarrier

\end{document}